\documentclass[%
superscriptaddress,
reprint,
showpacs,
 amsmath,amssymb,
 aps,
prb,
]{revtex4-1}

\newcommand\beq{\begin{equation}}
\newcommand\eeq{\end{equation}}

\usepackage{graphicx}
\usepackage{dcolumn}
\usepackage{bm}

\begin{document}


\title{Electromagnetic Tunneling through a Single-Negative Slab\\ Paired with a Double-Positive Bi-Layer}

\author{Giuseppe Castaldi}
\author{Ilaria Gallina}
\author{Vincenzo Galdi}
\email{vgaldi@unisannio.it}
 \homepage{http://www.ing.unisannio.it/vgaldi}
\affiliation{Waves Group, Department of Engineering, University of Sannio, I-82100 Benevento, Italy
}%

\author{Andrea Al\`u}
\affiliation{Department of Electrical and Computer Engineering, The University of Texas at Austin, Austin, TX 78712, USA}%

\author{Nader Engheta}
\affiliation{Department of Electrical and Systems Engineering, University of Pennsylvania, Philadelphia, PA 19104, USA}%

\date{\today}


\begin{abstract}
We show that resonant tunneling of electromagnetic fields can occur through a three-layer structure composed of a {\em single-negative} (i.e., either negative-permittivity or negative-permeability) slab paired with bi-layer made of {\em double-positive} (i.e., positive permittivity and permeability) media. In particular, one of the two double-positive media can be chosen arbitrarily (even vacuum), while the other may exhibit {\em extreme} (either near-zero or very high) permittivity/permeability values. Our results on this counterintuitive tunneling phenomenon also demonstrate the possibility of synthesizing double-positive slabs that effectively exhibit single-negative-like wave-impedance properties within a moderately wide frequency range.  
\end{abstract}

\maketitle
The past decade has witnessed a growing interest in the study and applications of {\em extraordinary} transmission of electromagnetic (EM) fields through {\em opaque} media, such as metals or, more in general, {\em single-negative} (SNG) materials characterized by {\em negative} permittivity or permeability, and hence {\em evanescent} fields. With reference to metallic films, surface-plasmon-induced transparency was demonstrated theoretically and experimentally in Ref. \onlinecite{Dragila}, and a large body of results are available in connection with extraordinary transmission through subwavelength apertures (see, e.g., Ref. \onlinecite{Vidal} for a recent review).
In connection with general SNG media, it was shown in Ref. \onlinecite{Alu1} that, in spite of their {\em inherent opacity}, slabs of (homogeneous, isotropic) epsilon-negative (ENG) and mu-negative (MNG) media can give rise to {\em resonant tunneling} phenomena (with total transmission and zero phase-delay) when paired as a bi-layer. Starting from this basic configuration, several extensions and generalizations have been investigated (see, e.g., Refs. \onlinecite{Pendry}--\onlinecite{Fang}).
 
Moreover, multi-layer metallo-dielectric structures have been extensively studied in connection with near-field sub-wavelength imaging (see, e.g., Refs. \onlinecite{Pendry2}--\onlinecite{Pendry3}), generating a growing interest in the study of resonant tunneling of evanescent waves in configurations featuring SNG media paired with {\em double-positive} (i.e., positive permittivty and permeability -- DPS) materials. 
Interestingly, while it is impossible to achieve perfect transparency by pairing a single SNG slab with an arbitrary DPS one, anomalous tunneling phenomena have been demonstrated in configurations featuring an SNG slab {\em sandwiched} between suitable impedance-matching DPS layers. \cite{Zhou2,Kim} In particular, the results in Ref. \onlinecite{Zhou2}, pertaining to a symmetrical DPS-ENG-DPS three-layer and experimentally validated at microwave frequencies, have been extended in Ref. \onlinecite{Hooper} to more general (e.g., optical, quantum-mechanical) tunnel barriers.
Against the above background, we propose here another counterintuitive resonant tunneling mechanism which entails pairing an SNG slab with a bi-layer made of DPS media. Unlike the configuration in Ref. \onlinecite{Zhou2}, where the SNG slab is symmetrically sandwiched between two identical DPS layers, in our case the DPS layers are different and paired at one side only. This also allows direct comparisons with configurations involving paired ENG-MNG media, with the DPS bi-layer playing the role of an {\em effective} SNG slab.
%
\begin{figure}
\begin{center}
\includegraphics [width=8cm]{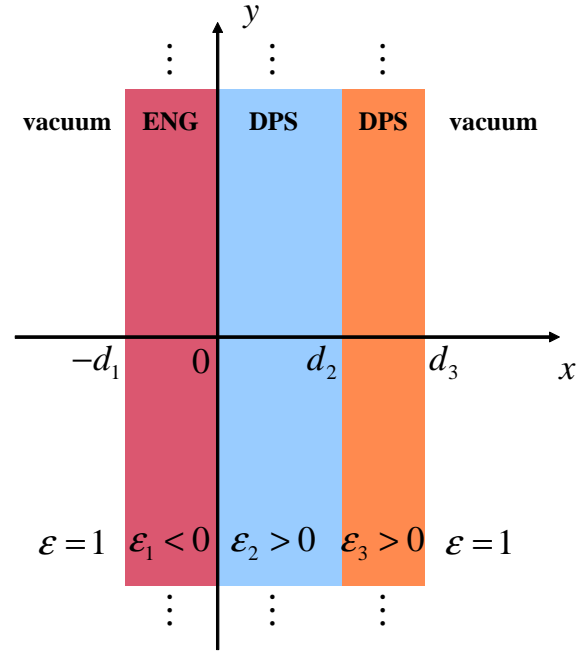}
\end{center}
\caption{(Color online) Problem geometry (details in the text).}
\label{Figure1}
\end{figure}

The problem geometry is illustrated in the Cartesian $(x,y,z)$ coordinate system of Fig. \ref{Figure1}.
Without loss of generality, we consider a homogeneous, isotropic ENG slab with relative permittivity $\varepsilon_1<0$ and thickness $d_1$ paired with a bi-layer composed of homogeneous, isotropic DPS media with parameters $\varepsilon_2>0$, $d_2$ and $\varepsilon_3>0$, $d_3$, respectively, all embedded in vacuum ($\varepsilon=1$) and assumed as {\em non-magnetic} (i.e., $\mu=1$),
under time-harmonic $[\exp(-i\omega t)]$ plane-wave illumination. 

Assuming normal incidence, for the ideal case of {\em lossless} media, we can derive via straightforward transfer-matrix algebra \cite{Born} the general resonance condition for {\em total transmission}, by zeroing the reflection coefficient, viz.,
\begin{eqnarray}
 &i&\left(1 - \varepsilon_1\right)
 \sqrt{\varepsilon_2 \varepsilon_3}
 \tau_1
 + i\left(1 - \varepsilon_2\right)
 \sqrt{-\varepsilon_1 \varepsilon_3} 
 \tau_2\nonumber\\ 
 &+&
 i\left(1 - \varepsilon_3\right)
 \sqrt{-\varepsilon_1 \varepsilon_2} \tau_3
 +\sqrt{\varepsilon_3} \left(\varepsilon_2 -\varepsilon_1\right)
 \tau_1
 \tau_2\nonumber\\
 &+& 
 \sqrt{\varepsilon_2}\left(\varepsilon_3 -\varepsilon_1\right)
 \tau_1\tau_3
 +\sqrt{-\varepsilon_1} 
 \left(\varepsilon_3 -\varepsilon_2\right)
 \tau_2\tau_3\nonumber\\
 &+& 
 i\left(\varepsilon_1 \varepsilon_3 -\varepsilon_2\right)
 \tau_1\tau_2\tau_3=0,
\label{eq:tot_TR}
\end{eqnarray}
where 
$\tau_1=\tanh\left(k
\sqrt{-\varepsilon_1}d_1
\right)$ and $\tau_{2,3}=\tan\left(
k\sqrt{\varepsilon_{2,3}}d_{2,3}
\right)$,
with $k=\omega/c=2\pi/\lambda$ denoting the vacuum wavenumber, and $c$ and $\lambda$ the corresponding wavespeed and wavelength. 
Zeroing the real part of (\ref{eq:tot_TR}), we obtain
\beq
\tau_2=
\frac{
\sqrt{\varepsilon_2} \left(\varepsilon_1-\varepsilon_3\right)
\tau_1
\tau_3}{{\sqrt{\varepsilon_3} \left({\varepsilon_2-\varepsilon_1}\right)
\tau_1 + \sqrt{-\varepsilon_1} \left({\varepsilon_3-\varepsilon_2}\right)\tau_3}},
\label{eq:tau2}
\eeq
which, enforced in the imaginary part of (\ref{eq:tot_TR}), yields
\beq
\tau_3= 
\pm 
\sqrt{\frac{{\varepsilon_3 \left(1 - \varepsilon_1\right)
\left(\varepsilon_2-\varepsilon_1\right)
\tau_1^2}}{{\varepsilon_1 \left(1 - \varepsilon_3\right)\left(\varepsilon_3-\varepsilon_2\right)
+\left(\varepsilon_3-\varepsilon_1\right)\left(\varepsilon_3 \varepsilon_1-\varepsilon_2\right)
\tau_1^2}}}. 
\label{eq:tau3}
\eeq
Noting that the numerator of the square-root argument in (\ref{eq:tau3}) is always positive (since $\varepsilon_1<0$ and $\varepsilon_{2,3}>0$), real solutions exist iff
\beq
\varepsilon _1 \left(1 - \varepsilon_3\right)\left(\varepsilon_3-\varepsilon_2\right)+\left(\varepsilon_3-\varepsilon_1\right)\left(\varepsilon_3 \varepsilon_1  - \varepsilon _2\right)\tau_1^2 > 0.
\label{eq:ineq1}
\eeq
The above inequality is quadratic in $\varepsilon_3$ (with positive discriminant), and yields the conditions
\beq
\varepsilon_3 \le\varepsilon_{3a},~~\mbox{or}~~\varepsilon_3 \ge\varepsilon_{3b},
\label{eq:ineq2}
\eeq
with $\varepsilon_{3a}$ and $\varepsilon_{3b}$ denoting the two (positive) roots of the denominator of the square-root argument in (\ref{eq:tau3}). It can be readily shown from (\ref{eq:ineq1}) that $\varepsilon_{3a}<\varepsilon_2$ and $\varepsilon_{3b}=\varepsilon_2/\varepsilon_{3a}>\varepsilon_2$,
from which it emerges the impossibility of achieving total transmission by pairing an ENG slab with a {\em single} DPS slab in vacuum, consistent with Ref. \onlinecite{Zhou2}.  Conversely, in the presence of a DPS {\em bi-layer}, for a given ENG slab with parameters $\varepsilon_1, d_1$, it is {\em always} possible [from (\ref{eq:tau2})--(\ref{eq:ineq2})] to find suitable parameters 
for the DPS layers so as to fulfill the total-transmission condition in (\ref{eq:tot_TR}) at a given frequency. More specifically, the parameter $\epsilon_2$ turns out to be {\em completely arbitrary} (but positive), while $\varepsilon_3$, $d_3$, and $d_2$ are derived proceeding backwards from (\ref{eq:ineq2}), (\ref{eq:tau3}), and (\ref{eq:tau2}), respectively. 

A few general considerations are in order. First, at variance with the symmetrical DPS-ENG-DPS configuration in Ref. \onlinecite{Zhou2}, {\em no} critical thickness value exists for our asymmetrical ENG-DPS-DPS configuration in order to exhibit EM tunneling. In fact, there always exist {\em four} distinct classes of solutions for total transmission, corresponding to the possible combinations of the choices in (\ref{eq:tau3}) and (\ref{eq:ineq2}). We found that, in the long-wavelength limit $d_{n}/\lambda_0\rightarrow 0$, $n=1,2,3$, {\em all} these solutions tend towards the effective-medium-theory prediction obtained from (\ref{eq:tot_TR}) by neglecting the multiple-scattering terms (i.e., those containing multiple $\tau$-terms), 
\beq
\varepsilon_1d_1+\varepsilon_2d_2+\varepsilon_3d_3=d_1+d_2+d_3.
\label{eq:EMT}
\eeq
It can be shown that, for an increasing opacity of the ENG medium ($|\varepsilon_1|\gg1$), the end values in (\ref{eq:ineq2}) tend to exhibit {\em extreme values}, so that the outermost DPS slab must have either {\em near-zero} ($\varepsilon_{3a}\ll 1$) or {\em very-high} ($\varepsilon_{3b}\gg1$) permittivity. From (\ref{eq:tau3}), it is readily verified that the solutions associated to the end values $\varepsilon_3=\varepsilon_{3a}$ or $\varepsilon_3=\varepsilon_{3b}$ in (\ref{eq:ineq2}) correspond to a {\em quarter-wavelength} (plus half-wavelength periodicities) size for the outermost DPS slab, which (recalling its wave-impedance-transformation properties \cite{Born}) is also representative of a different configuration featuring an ENG-DPS bi-layer terminated with a DPS half-space (i.e., $d_3\rightarrow\infty$).  
Moreover, we highlight that the {\em complete arbitrariness} in the choice of $\varepsilon_2$ represents an important degree of freedom in the proposed configuration. In particular, choosing $\varepsilon_2=1$, we obtain another interesting tunneling configuration featuring an ENG slab paired with a DPS slab via a separating vacuum layer.
Finally, it can be shown that enforcing a {\em purely real} (i.e., $\pm1$) transmission coefficient results in 
\begin{eqnarray}
\left({\varepsilon_1^2  - \varepsilon_1 \varepsilon_2  - \varepsilon_1 \varepsilon_3  + \varepsilon_2 \varepsilon_3}\right)^2  = \left({1 - \varepsilon_1 }\right)\left( {\varepsilon_2  - \varepsilon _1 } \right)\nonumber\\
\times\!\left[ {\frac{\varepsilon_1}{\tau_1^2} 
\!\left(\varepsilon_3  \!-\! \varepsilon_2\right)\left(\varepsilon _3 \!-\! 1\right)
\!+\! \left( {\varepsilon _1  \!-\! \varepsilon _3 } \right)\left(\varepsilon_1 \varepsilon_3 \!-\!\varepsilon_2\right)} \right],
\end{eqnarray}
i.e., a quadratic equation in $\varepsilon_3$, whose solutions, enforced in (\ref{eq:ineq1}), 
yield the constraint
\beq
{\varepsilon _1^2 \left( {1 - \varepsilon _1 } \right)\left( {\varepsilon _1  - \varepsilon _2 } \right)^3 \tau _1^2 }>0,
\eeq
which is clearly {\em impossible} to fulfill with $\varepsilon_1<0$ and $\varepsilon_2>0$, thereby implying that no tunneling with zero phase-delay is possible in our proposed configuration.
%
\begin{figure}
\begin{center}
\includegraphics [width=8cm]{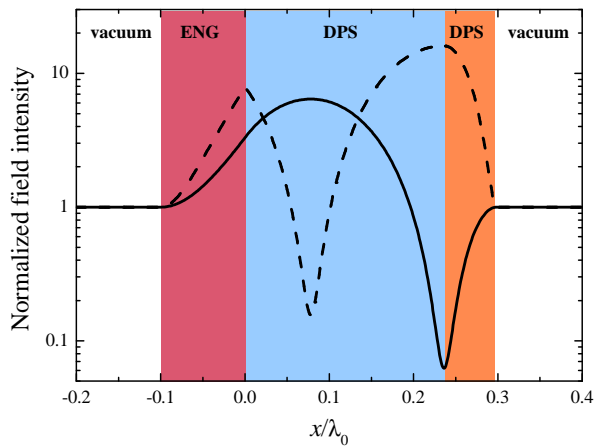}
\end{center}
\caption{(Color online) Resonant electric (solid) and magnetic (dashed) field intensity distributions (normalized by the incident fields) for a configuration as in Fig. \ref{Figure1}, with $\varepsilon_1=-3$, $d_1=0.1\lambda_0$, $\varepsilon_2=2.5$, $d_2=0.236\lambda_0$, $\varepsilon_3=16$, $d_3=0.0624\lambda_0$.}
\label{Figure2}
\end{figure}

As a first example, we consider an ideal {\em lossless} configuration featuring an ENG slab with $\varepsilon_1=-3$ and $d_1=0.1\lambda_0$ (here and henceforth, the subscript $_0$ identifies resonant frequency/wavelength quantities). We arbitrarily select $\varepsilon_2=2.5$, and readily derive from (\ref{eq:tau2})--(\ref{eq:ineq2}) the remaining parameters: $\varepsilon_3=\varepsilon_{3b}=16$, $d_3=0.0624\lambda_0$ [choosing the positive determination for $\tau_3$ in (\ref{eq:tau3})], and $d_2=0.236\lambda_0$.
For a normally-incident (along the positive $x$-direction) plane-wave excitation, Fig. \ref{Figure2} illustrates the resonant electric and magnetic field (normalized) longitudinal distributions, from which it can be observed the total-transmission effect, achieved via a growing evanescent wave in the ENG layer and a standing wave in the DPS bi-layer. The electric field peaks towards the center of the middle DPS layer, and the magnetic field peaks at its boundaries, similar to a Fabry-Perot etalon, although we consider here the presence of an inherently opaque material.
%
\begin{figure}
\begin{center}
\includegraphics [width=8cm]{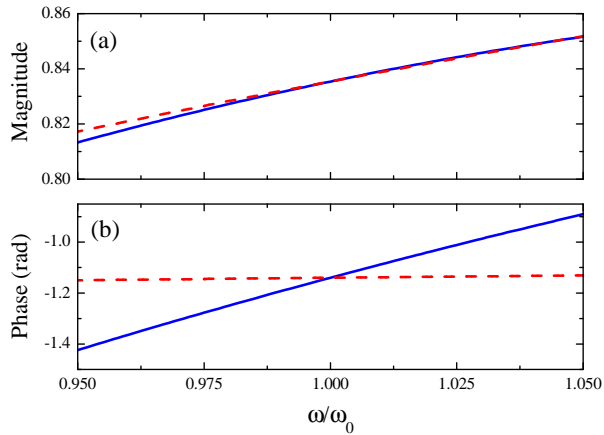}
\end{center}
\caption{(Color online) Reflection coefficient magnitude (a) and phase (b) of the isolated DPS bi-layer in Fig. \ref{Figure2} (blue-solid) as a function of the frequency, compared with the response (red-dashed) of an effective MNG slab with $\varepsilon_e=1$, $\mu_e=-0.33$.}
\label{Figure3}
\end{figure}
It is insightful to compare the above phenomenon with the tunneling occurring in an ENG-MNG pair. From the ENG-MNG matching conditions [cf. Eq. (8) in Ref. \onlinecite{Alu1}], we can straightforwardly derive the constitutive parameters $\varepsilon_e=1$ and $\mu_e=-0.33$ of the required MNG slab of same thickness ($d_2+d_3$) as the above DPS bi-layer to ``compensate'' the given ENG slab. Figure \ref{Figure3} compares the reflection-coefficient responses of the {\em isolated} DPS bi-layer and such MNG slab (free-standing in vacuum) as a function of frequency. As expected, both responses result in large reflectivity (due to the opacity of an MNG stand-alone layer) and {\em perfectly match} at the resonance frequency. In particular, the designed DPS bi-layer is capable of providing the capacitive input impedance required to resonate with the ENG slab, ensuring total transmission. The reflection coefficients in Fig. \ref{Figure3} agree reasonably well within a moderate bandwidth (maximum difference of $\sim 0.5\%$ in magnitude and $\sim \pi/12$ in phase, over a $10\%$ bandwidth), ensuring that the designed DPS bi-layer may effectively replace an MNG layer for a variety of applications.

%
\begin{figure}
\begin{center}
\includegraphics [width=8.cm]{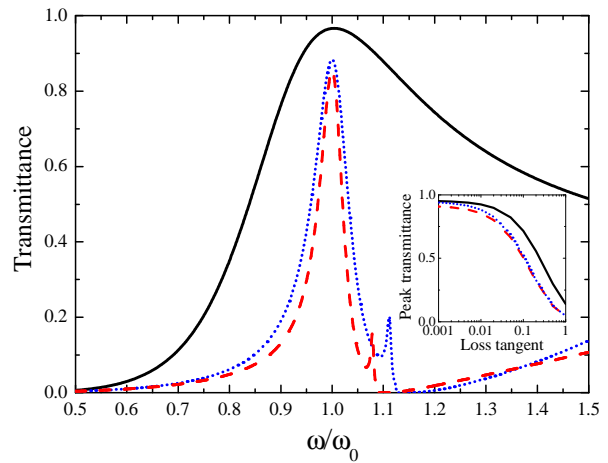}
\end{center}
\caption{(Color online) Black-solid: Transmittance as a function of the frequency for the parameter configuration in Fig. \ref{Figure2}, but using for the ENG slab the dispersive model in (\ref{eq:Drude}) with $\omega_{p1}=2\omega_0$, $\gamma_1=3.75\cdot10^{-3}\omega_{p1}$ (i.e., $\mbox{Re}[\varepsilon_3(\omega_0)]\approx-3$), and a loss-tangent of $10^{-3}$ for the DPS layers. Also shown are the responses obtained for an increased opacity of the ENG slab ($\omega_{p1}=10.05\omega_0$ and $\gamma_1=9.8\cdot10^{-4}\omega_{p1}$ in (\ref{eq:Drude}), i.e., $\mbox{Re}[\epsilon_1(\omega_0)]\approx-100$, and $d_1=0.01\lambda_0$), for $\varepsilon_2=12$ (with loss-tangent=$10^{-3}$) and $d_2=0.119\lambda_0$ (red-dashed), and $\varepsilon_2=1$ and $d_2=0.472\lambda_0$ (blue-dotted), with the high-permittivity outermost DPS layer described by the model in (\ref{eq:Lorentz}), with $\varepsilon_{3\infty}=4$, $\Lambda_3=3.081\omega_3$, and $\omega_3=1.091\omega_0$, $\gamma_3=9.26\cdot 10^{-4}\omega_3$ (i.e., $\mbox{Re}[\varepsilon_3(\omega_0)]=63.7$), $d_3=0.036\lambda_0$, and $\omega_3=1.13\omega_0$, $\gamma_3=1.33\cdot 10^{-3}\omega_3$ (i.e., $\mbox{Re}[\varepsilon_3(\omega_0)]=47.8$), $d_3=0.0313\lambda_0$, respectively. The inset illustrates the influence of losses in the ENG material on the peak transmittance (as a function of the loss-tangent at resonance).}
\label{Figure4}
\end{figure}

%
\begin{figure}
\begin{center}
\includegraphics [width=8cm]{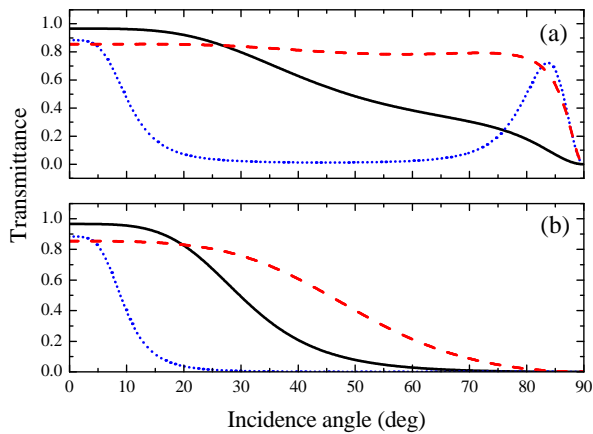}
\end{center}
\caption{(Color online) As in Fig. \ref{Figure4}, but as a function of the incidence angle at the resonant frequency, for the P (a) and S (b) polarizations.}
\label{Figure5}
\end{figure}

Next, we move on to assessing the frequency and angular dependence, as well as the sensitivity to polarization and losses, of the above tunneling phenomenon. Figures \ref{Figure4} and \ref{Figure5} show the transmittance as a function of frequency and incidence angle (from the $x-$axis, for both P and S polarizations), respectively, for three representative parameter configurations. We start from the same configuration in Fig. \ref{Figure2}, but now assuming a more realistic Drude-type dispersive, lossy model for the ENG medium,
\beq
\varepsilon_1\left(\omega\right)=1-\frac{\omega_{p1}^2}{\omega\left(\omega+i\gamma_1\right)},
\label{eq:Drude}
\eeq
with the plasma angular frequency $\omega_{p1}$ and the damping coefficient $\gamma_1$ chosen so as to ensure $\mbox{Re}[\varepsilon_1(\omega_0)]\approx-3$ (with a loss-tangent $\sim 10^{-2}$), and a non-dispersive slightly-lossy (loss-tangent=$10^{-3}$) model for the DPS layers.  From Fig. \ref{Figure4} (black-solid curve), a rather broad resonance is observed in this case, with a peak-transmittance of nearly $97\%$, and a slow decay for higher frequencies attributable to the increasing transparency (approaching the plasma frequency) of the ENG slab. Also the angular response (see Fig. \ref{Figure5}) turns out not to be very selective, especially for P-polarization.
As previously highlighted, tunneling effects may be obtained, in principle, for {\em arbitrary} choices of the ENG slab parameters and the permittivity of the middle DPS layer. For an increased opacity of the ENG medium, as anticipated, the constitutive parameters of the outermost DPS layer tend to exhibit extreme values. For instance, considering an ENG slab with $\varepsilon_1=-100$ and $d_1=0.01\lambda_0$, and $\varepsilon_2=12$, we obtain [choosing the larger end-value $\varepsilon_{3b}$ in (\ref{eq:ineq2})] $\varepsilon_3=63.7$. In Ref. \onlinecite{Zhou2}, such (positive/negative) high-permittivity media were successfully synthesized at microwave frequencies with resonant inclusions. Assuming for the outermost DPS medium a Lorentz-type dispersive, lossy model, 
\beq
\varepsilon_3\left(\omega\right)=
\varepsilon_{3\infty}-\frac{\Lambda_3^2}{\omega^2-\omega_3^2+2i\gamma_3\omega},
\label{eq:Lorentz}
\eeq
with the parameters (given in Fig. \ref{Figure4} caption) tuned so as to ensure the required real-part at the given frequency, we observe in Fig. \ref{Figure4} (red-dashed curve) a narrower bandwidth (with peak-transmittance of nearly $85 \%$ due to the larger sensitivity to losses, and a smaller transmission peak attributable to dispersion effects), and from Fig. \ref{Figure5} a {\em flatter} angular response (especially for P-polarization) which is a direct consequence of the increased permittivity values. \cite{Zhou2} The resonant field distributions, not shown here for brevity, are qualitatively similar to those in Fig. \ref{Figure2}, with intensity enhancements of nearly a factor 60. Also shown in Figs. \ref{Figure4} and \ref{Figure5} (blue-dotted curves) are the responses obtained for the same ENG slab, but choosing $\varepsilon_2=1$, i.e., an ENG-vacuum-DPS configuration. While the frequency response and the field distributions (again, not shown for brevity) resemble the previous example, the angular response is now much more selective (for both polarizations), and exhibits (in the P-polarization case) another peak for near-grazing incidence, which corresponds to a pseudo-Brewster angle of the ENG slab.   
The sensitivity to losses in the ENG material is illustrated in the inset of Fig. \ref{Figure4}, in terms of the peak transmittance as a function of the loss-tangent (at resonance), and qualitatively resembles the behavior observed in other tunneling phenomena involving SNG materials. \cite{Alu1} Finally, our simulations show that, for given parameters of the ENG slab, the sensitivity to variations within $\sim 1/25$ of the local wavelength with respect to the nominal DPS slab thicknesses $d_{2,3}$ would still yield moderate reductions ($\sim -3$dB) of the peak transmittance (with possible frequency shifts).

To sum up, we have illustrated a counterintuitive EM-field resonant tunneling mechanism that can take place by pairing an SNG slab with a DPS bi-layer. In particular, we have worked out analytically the design criteria, and explored the dependence on frequency, incidence direction, polarization, and material losses. Our results, which can be extended (along the lines of Refs. \onlinecite{Hooper} and \onlinecite{Jelinek}) to more general (e.g., optical, quantum-mechanical) {\em asymmetrical} tunnel barriers, also allow direct comparisons with ENG-MNG paired configurations,   
and can be understood in terms of the equivalent wave-impedance properties exhibited by a DPS bi-layer and a ``matched'' (according to Ref. \onlinecite{Alu1}) SNG slab. This observation may suggest more general application scenarios of such DPS bi-layer, for which SNG-like responses may be {\em emulated} via simpler dielectric slabs, which may somehow compensate, in a simple geometry, opaque ENG metamaterial slabs for a variety of applications in which {\em complementary} metamaterials are paired together. \cite{Alu1,Pendry} Also of interest is the possible adaptation for applications to super-resolution imaging schemes.

\end{document}